\documentclass[conference]{IEEEtran}
\IEEEoverridecommandlockouts
\usepackage{cite}
\usepackage{amsmath,amssymb,amsfonts}
\usepackage{algorithmic}
\usepackage{graphicx}
\usepackage{textcomp}
\usepackage{xcolor}
\def\BibTeX{{\rm B\kern-.05em{\sc i\kern-.025em b}\kern-.08em
    T\kern-.1667em\lower.7ex\hbox{E}\kern-.125emX}}

\usepackage[linesnumbered,ruled,vlined]{algorithm2e}
\usepackage[hyphens]{url}
\usepackage{subcaption}
\usepackage{ctable} % for \specialrule command
\usepackage{makecell}
\usepackage{multirow}
\usepackage{array}
\newcolumntype{P}[1]{>{\centering\arraybackslash}p{#1}}
\newcolumntype{M}[1]{>{\centering\arraybackslash}m{#1}}
\begin{document}

\title{RACE-IT: A Reconfigurable Analog Computing
Engine for In-Memory Transformer Acceleration}

\author{
\IEEEauthorblockN{
Lei Zhao\textsuperscript{1},
Aishwarya Natarajan\textsuperscript{1},
Luca Buonanno\textsuperscript{1},
Archit Gajjar\textsuperscript{1},
Ron Roth\textsuperscript{2}, \\
Sergey Serebryakov\textsuperscript{1},
John Moon\textsuperscript{1},
Omar Eldash\textsuperscript{1},
Jim Ignowski\textsuperscript{1},
Giacomo Pedretti\textsuperscript{1}
}

\vspace{1em}

\IEEEauthorblockA{
\textsuperscript{1}Artificial Intelligence Research Lab (AIRL), Hewlett Packard Labs, USA\\
\textsuperscript{2}Technion - Israel Institute of Technology, Haifa, Israel
}

\vspace{0.5em}

\IEEEauthorblockA{
Emails: \{lei.zhao, aishwarya.natarajan, luca.buonanno, archit.gajjar, sergey.serebryakov, \\jmoon, oeldash, jim.ignowski, giacomo.pedretti\}@hpe.com, ronny@cs.technion.ac.il
}
}

%\and
%\and
%\and
%\and
%\and

\maketitle

\begin{abstract}
Transformer models represent the cutting edge of Deep Neural Networks (DNNs) and excel in a wide range of machine learning tasks.
However, processing these models demands significant computational resources and results in a substantial memory footprint.
While In-memory Computing (IMC) offers promise for accelerating Vector-Matrix Multiplications (VMMs) with high computational parallelism and minimal data movement, employing it for other crucial DNN operators remains a formidable task.
This challenge is exacerbated by the extensive use of complex activation functions, Softmax, and data-dependent matrix multiplications (DMMuls) within Transformer models.

To address this challenge, we introduce a Reconfigurable Analog Computing Engine (RACE) by enhancing Analog Content Addressable Memories (ACAMs) to support broader operations.
Based on the RACE, we propose the RACE-IT accelerator (meaning RACE for In-memory Transformers) to enable efficient analog-domain execution of all core operations of Transformer models.
Given the flexibility of our proposed RACE in supporting arbitrary computations, RACE-IT is well-suited for adapting to emerging and non-traditional DNN architectures without requiring hardware modifications.

We compare RACE-IT with various accelerators.
Results show that RACE-IT increases performance by 453$\times$ and 15$\times$, and reduces energy by 354$\times$ and 122$\times$ over the state-of-the-art GPUs and existing Transformer-specific IMC accelerators, respectively.
\end{abstract}

\begin{IEEEkeywords}
in-memory computing, analog computing, neural network, RRAM
\end{IEEEkeywords}

\section{Introduction}
The Transformer architecture~\cite{vaswani2017attention} has gained immense popularity in the fields of natural language processing~\cite{Lan2020ALBERT}, computer vision~\cite{li2022exploiting}, etc., thanks to its remarkable ability to capture complex patterns in sequential and image data.
However, the performance of Transformers comes at a cost: their high demand for computational resources and memory.
These models often feature billions, if not trillions, of parameters, demanding substantial computation and data movement.
The sheer scale of Transformer models, exemplified by LLaMA \cite{grattafiori2024llama} and DeepSeek \cite{liu2024deepseek}, necessitates specialized hardware and high-capacity memory systems, presenting a significant challenge to deploy them at scale.
As a result, efficiently processing Transformer models remains a crucial concern in the quest to unlock their full potential for a wider range of applications.

In-memory Computing (IMC) stands out as a promising solution to alleviate the computational and memory challenges posed by Transformer models \cite{yang2020retransformer, kang2021framework, zheng2023accelerating}.
IMC involves performing computations directly within the memory, bypassing the need for frequent data movement between memory and the processing units, which is a significant bottleneck in traditional computing architectures.
Emerging memories such as Resistive Random Access Memories (RRAM) and Phase Change Memories (PCMs) offer a compelling avenue for IMC due to their non-volatile nature and high integration potential.
Crossbar arrays of such emerging memories can efficiently map neural network weights in their conductance and accelerate a layer evaluation by parallel Vector-Matrix Multiplications (VMMs) in the analog domain \cite{hu2016dac}.
High-performance silicon demonstrations of multi-core IMC accelerators using RRAM~\cite{wan_compute--memory_2022} and PCM~\cite{le_gallo_64-core_2023} have been recently presented, finally grounding years of research in experimental measurements.
However, one notable drawback is that these designs may not be optimally suited for critical operations like activations, Softmax, and Data-dependent Matrix Multiplications (DMMul), which are fundamental in the Transformer architecture.
%Addressing these limitations and optimizing IMC designs for these specific operations is a critical area of ongoing research.% to enhance the applicability of IMC in Transformer-based processing.

To address this challenge, existing IMC designs typically follow one of three main approaches.
The first is an ad-hoc strategy that integrates specialized CMOS-based units tailored to specific workloads, as exemplified by ISAAC~\cite{shafiee2016isaac}.
While effective for fixed functionality, this approach lacks the flexibility to support new operations beyond Convolutional Neural Networks (CNNs).
The second approach, adopted by architectures such as PUMA~\cite{ankit2019puma}, introduces programmable digital Vector Functional Units (VFUs) to support a wide range of operations. 
However, this programmability comes at the cost of performance, particularly for DMMuls in Transformer models.
The third approach, as demonstrated in ReTransformer~\cite{yang2020retransformer}, directly computes DMMuls and Softmax within RRAM crossbars by programming the inputs into the array.
While promising, it suffers from the inherent drawbacks of RRAM's slow writes and limited endurance~\cite{ielmini2025resistive}.
Moreover, operations such as the GeLU activation remain unsupported in the analog domain and require dedicated digital hardware.
Table~\ref{tab:comparison} shows the comparison of these approaches with our proposed design.

\begin{table}[h!]
\centering
\caption{Analog IMC approaches.}
\label{tab:comparison}
\scriptsize
\begin{tabular}{c|M{0.3in}M{0.3in}M{0.4in}M{0.3in}M{0.4in}}
\textbf{Architecture} & \textbf{VMM} & \textbf{Act} & \textbf{Softmax} & \textbf{DMMul} & \textbf{Support Transformers} \\
\specialrule{.2em}{.1em}{.1em}
ISAAC \cite{shafiee2016isaac} & Crossbar & Fixed logic & - & - & {\color{purple}\textbf{No}}\\
\hline
PUMA \cite{ankit2019puma} & Crossbar & VFU & VFU & VFU & {\color{teal}\textbf{Yes}}\\
\hline
ReTransformer \cite{yang2020retransformer} & Crossbar & VFU & Crossbar & Crossbar & {\color{teal}\textbf{Yes}}\\
\hline
RACE-IT (this work) & Crossbar & RACE & RACE & RACE & {\color{teal}\textbf{Yes}}\\
\hline
\end{tabular}
\end{table}

In this paper, we present a novel Reconfigurable Analog Computing Engine (RACE) built upon RRAM-based Analog Content Addressable Memory (ACAM)~\cite{li2020analog} to support a wide range of arbitrary operations.
By reprogramming the value ranges stored in RRAM devices, the RACE can be reconfigured to perform different computations without requiring any hardware modifications.
This reconfigurability makes it a compelling solution for building architectures capable of supporting all core operations in Transformer models, as well as future operators yet to be conceived.
To reduce hardware overhead, we introduce an encoding scheme that significantly reduces the size of RACE arrays, thus further improving the area and energy efficiency.

While the RACE offers energy-efficient computation by operating in the analog domain, it inherits the susceptibility to analog noise and device non-idealities.
To evaluate its real-world effectiveness, we perform detailed circuit simulations in SPICE and propose a noise-aware fine-tuning (NAF) strategy by incorporating realistic noise models and variability to significantly improve CNNs and Transformer model's inference robustness under practical operating conditions.

By integrating the RACE with RRAM crossbars to form an IMC system, our proposed RACE-IT (i.e., RACE for In-memory Transformers) accelerator demonstrates significant performance improvements and energy savings over state-of-the-art GPUs, systolic arrays, and prior IMC designs.%—achieving up to 453$\times$ speedup, 354$\times$ energy saving over GPUs, and 15$\times$ speedup, 122$\times$ energy saving over existing IMC accelerators, respectively.

In summary, this paper makes the following contributions:

\begin{itemize}
\item We propose the RACE, a compact and reconfigurable analog structure that supports arbitrary operations, enabling flexible analog DNN acceleration.

\item We introduce an encoding-based optimization that reduces the RACE array size and significantly lowers hardware overhead.

\item We implement detailed circuit simulations of the RACE. To mitigate analog noise, we develop a NAF method that enhances inference robustness in IMC systems.

\item We conduct comprehensive comparisons against state-of-the-art GPUs, systolic array accelerators, and IMC designs employing alternative methods for non-VMM operations in Transformers. Experimental results demonstrate substantial performance improvements and energy savings.
\end{itemize}

\section{Background}

\subsection{Crossbar-based VMM Computing Engine}

\begin{figure}[htbp]
\centerline{\includegraphics[width=1.8in]{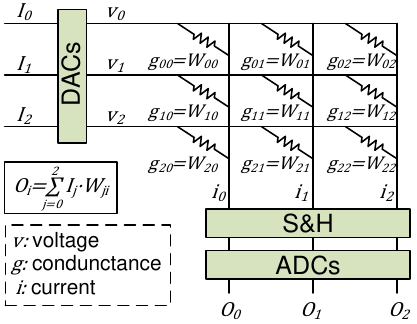}}
\caption{RRAM crossbar to compute VMM.}
\label{fig:crossbar}
\end{figure}

RRAM is an emerging non-volatile memory technology with strong potential for accelerating VMMs when arranged in crossbar arrays~\cite{ielmini2025resistive}.
Figure~\ref{fig:crossbar} illustrates a VMM example where a 3-element input feature vector ($I$) is multiplied by a $3 \times 3$ weight matrix ($W$) using a RRAM crossbar.
Digital-to-Analog Converters (DACs) convert input elements into a voltage vector ($v$), which is applied to the word lines.
Weights are stored as RRAM conductance values ($g$), and Kirchhoff's law ensures that the resulting current ($i$) on each bit line reflects the dot product of the input voltages and the conductance values.
These analog currents are sampled by Sample-and-Hold (S\&H) units and then digitized by Analog-to-Digital Converters (ADCs) to produce the digital output vector ($O$).
While the figure shows a small $3 \times 3$ crossbar for clarity, practical systems often use multiple much larger arrays (e.g., 512×256~\cite{chen201865nm}) to exploit massive parallelism.

\subsection{Analog Content Addressable Memory}
\begin{figure}[htbp]
\vspace{-0.1in}
\centerline{\includegraphics[width=3.3in]{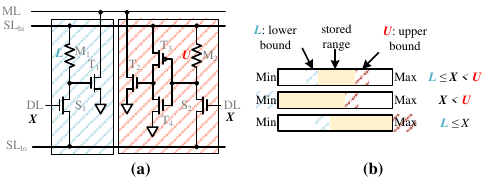}}
\caption{(a) An ACAM cell structure. (b) The operational principle of an ACAM cell.}
\label{fig:acam}
\vspace{-0.1in}
\end{figure}
Content Addressable Memories (CAMs) are a memory structure ubiquitous in networking, in which an input data vector is searched in the memory and its location is returned as output~\cite{pagiamtzis_content-addressable_2006}.
%A conventional digital CAM cell is limited to comparing a single input binary bit with a stored binary bit.
%In Ternary CAMs (TCAMs), a wildcard can be used both as input and as a stored value to match either 1s or 0s.
The recently proposed ACAM cell structure~\cite{li2020analog} allows comparing an analog input value within a defined range stored in the cell.
Figure~\ref{fig:acam}(a) presents the schematic of an ACAM cell.
Two RRAMs, $M_1$ and $M_2$, are utilized to store the lower and upper bounds of a range, respectively.
The blue-shaded component compares the input $X$ on the Data Line ($DL$) with the lower bound $L$ of the range, while the red-shaded component compares the input $X$ with the upper bound $U$ of the range.
The combined operation of the series transistors ($S_1$ and $S_2$) and RRAMs acts as voltage dividers, controlling the voltage supplied to the pull-down transistors ($T_1$ and $T_2$).
In the blue-shaded component, when the input is greater than or equal to the data stored in $M_1$, $T_1$ is off, thus keeping the Matching Line ($M$L) at a high voltage.
The red-shaded component operates similarly, with an inverter ($T_3$ and $T_4$) added before the pull-down transistor $T_2$.
Figure~\ref{fig:acam}(b) illustrates the operational principle of an ACAM cell.
It is possible to represent a "Don't Care" value on either side of the range by programming $M_1$ ($M_2$) to the lowest (highest) possible conductance, as depicted by the lower two rows in Figure~\ref{fig:acam}(b).
Studies have demonstrated up to 4-bit analog search in ACAM~\cite{li2020analog, pedretti2021tree}.

\subsection{Transformers and Multi-Head Attention}

\begin{figure}[htbp]
\vspace{-0.2in}
\centerline{\includegraphics[width=1in]{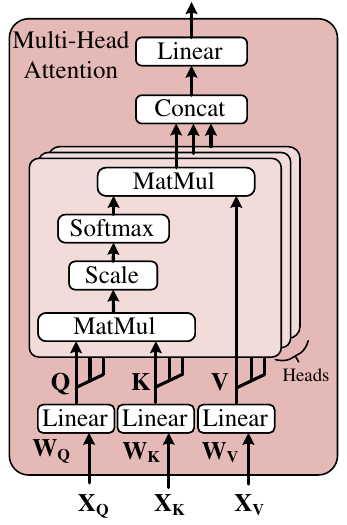}}
\vspace{-0.1in}
\caption{Multi-Head attention in Transformers.}
\label{fig:transformer}
\vspace{-0.1in}
\end{figure}

The most fundamental component of Transformer models is the Multi-Head Attention (MHA), depicted in Figure~\ref{fig:transformer}.
MHA operates on a sequence of tokens as its input, comprising three types of token sequences: $\mathbf{x_Q}$, $\mathbf{x_K}$, and $\mathbf{x_V}$.
For each of these token sequences, a distinct linear layer (denoted by $\mathbf{W_Q}$, $\mathbf{W_K}$, and $\mathbf{W_V}$) is applied to generate respective matrices: $\mathbf{Q}$, $\mathbf{K}$, and $\mathbf{V}$.
This operation is defined as:
\begin{equation}
\mathbf{Q}, \mathbf{K}, \mathbf{V} = \mathbf{x_Q} \cdot \mathbf{W_Q}, \mathbf{x_K} \cdot \mathbf{W_K}, \mathbf{x_V} \cdot \mathbf{W_V}
\end{equation}
These resulting matrices are then partitioned into multiple segments, often referred to as \textit{heads}, which undergo identical sets of computations, specifically the attention mechanism:
\begin{equation}
Attention(\mathbf{K_i},\mathbf{Q_i},\mathbf{V_i}) = \text{softmax}\left(\frac{\mathbf{Q_i} \cdot \mathbf{K_i}^T}{\sqrt{d_k}}\right) \cdot \mathbf{V_i}
\label{eq:attention}
\end{equation}
where $i \in num\_heads$, and $d_k$ represents the dimension of the query and key vectors.
Ultimately, the output from each head is concatenated and goes through another linear layer to form the overall output of this MHA.

\section{Reconfigurable Analog Computing Engine}

\begin{figure}[htbp]
\centerline{\includegraphics[width=3.3in]{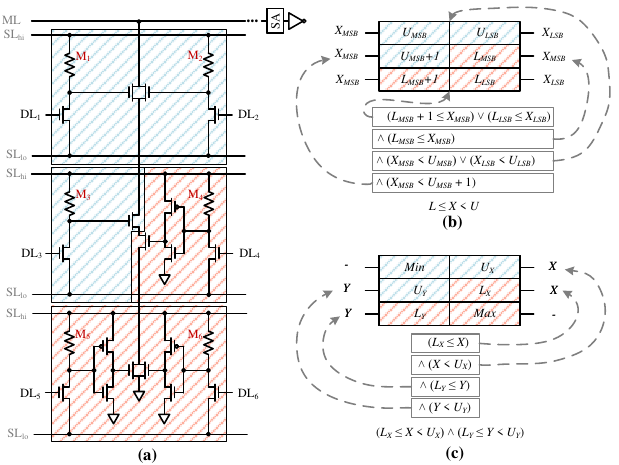}}
\caption{(a) A RACE cell structure. (b) Mapping a one-variable 8-bit comparison onto a RACE cell. (c) Mapping a two-variable 4-bit comparison onto a RACE cell.}
\label{fig:race}
\end{figure}

\subsection{Cell Structure}

Our goal is to design a cell capable of computing arbitrary one- and two-variable functions by decomposing them into single- and two-variable comparisons (see Section~\ref{sec:compute}).
To enable this, we propose a new RACE cell structure that supports both comparison types.
However, since each RRAM device in ACAM stores up to 4 bits, we aim to design a cell that compares partial bit widths and combines the results to support full 8-bit comparisons (see Section~\ref{sec:mode}).

Figure~\ref{fig:race}(a) shows the proposed RACE cell structure.
Recall that an ACAM cell in Figure~\ref{fig:acam}(a) consists of a blue component that holds the lower bound of the stored range, and a red component that holds the upper bound.
A single RACE cell comprises three blue components and three red components.
All six components are connected to the pull-down path of the $ML$.
The top two blue components are connected in parallel, implementing a logical AND operation: the $ML$ is pulled down to ground when both $(DL_1>M_1)$ $and$ $(DL_2>M_2)$.
Similarly, the bottom two red components are also connected in parallel, so the $ML$ is pulled down when both $(DL_5<M_5)$ $and$ $(DL_6<M_6)$.
These two AND combinations are then connected in series with the serially connected blue and red components at the center, which implement the conditions $(DL_3>M_3)$ $OR$ $(DL_4<M_4)$.
As a result, the overall operation performed by a RACE cell can be expressed as:

{\small
\vspace{-0.1in}
\begin{align}
\overline{ML}=&[(DL_1>M_1)\wedge(DL_2>M_2)] \nonumber\\
&\vee(DL_3>M_3) \nonumber\\
&\vee(DL_4<M_4) \nonumber\\
&\vee[(DL_5<M_5)\wedge(DL_6<M_6)] \nonumber
\end{align}
}

The inversion on $ML$ reflects the inverter placed after the sense amplifier (SA) in the RACE readout path.
By applying De Morgan's laws to compute inversion on both sides of the equation, we get:

{\small
\vspace{-0.1in}
\begin{align}
ML=&[(DL_1<M_1)\vee(DL_2<M_2)] \nonumber\\
&\wedge(DL_3<M_3) \nonumber\\
&\wedge(DL_4>M_4) \nonumber\\
&\wedge[(DL_5>M_5)\vee(DL_6>M_6)]
\end{align}
}

That is, the blue components store the upper bounds and the red components store the lower bounds.

\subsection{Computation Modes}
\label{sec:mode}

A RACE cell can be configured into two modes, according to the values stored in RRAM devices and the input on $DL$s.

\textbf{One-variable 8-bit comparison mode.}
In this mode, a RACE cell computes whether an 8-bit input value $X$ is inside an 8-bit range ($L<X<U$).
Although lower-bit quantization has been studied, 8-bit activation remains the most popular—especially in Transformer-based large language models (LLMs)—due to its balance of accuracy and ease of implementation~\cite{lin2024awq}.
However, RRAM devices in ACAMs can only store up to 4 bits.
We slice both the 8-bit input value and the bounds of the range into a 4-bit MSB part and a 4-bit LSB part, similar to \cite{pedretti2023x}.
The three blue components collectively represent the 8-bit upper bound of the range, and the three red components collectively store the 8-bit lower bound of the range.
Therefore, the 8-bit comparison of $L<X<U$ can be decomposed into a form shown in Figure~\ref{fig:race}(b).
The dotted lines show how the sub-clauses of the decomposed comparison are mapped to each component of the RACE cell.

\textbf{Two-variable 4-bit comparison mode.}
This mode computes whether two 4-bit inputs ($X$ and $Y$) are both inside their 4-bit ranges, i.e., $(L_x<X<U_x) \wedge (L_y<Y<U_y)$.
The decomposition and mapping of this computation is shown in Figure~\ref{fig:race}(c).

Note that the mode of the RACE is determined solely by the values stored in each cell component, without requiring any hardware modifications.
In the next section, we demonstrate how these two computation modes can be used to implement arbitrary functions.

\section{RACE-Based Computing}
\label{sec:compute}

\subsection{Activation}

\begin{figure}[htbp]
\centerline{\includegraphics[width=3in]{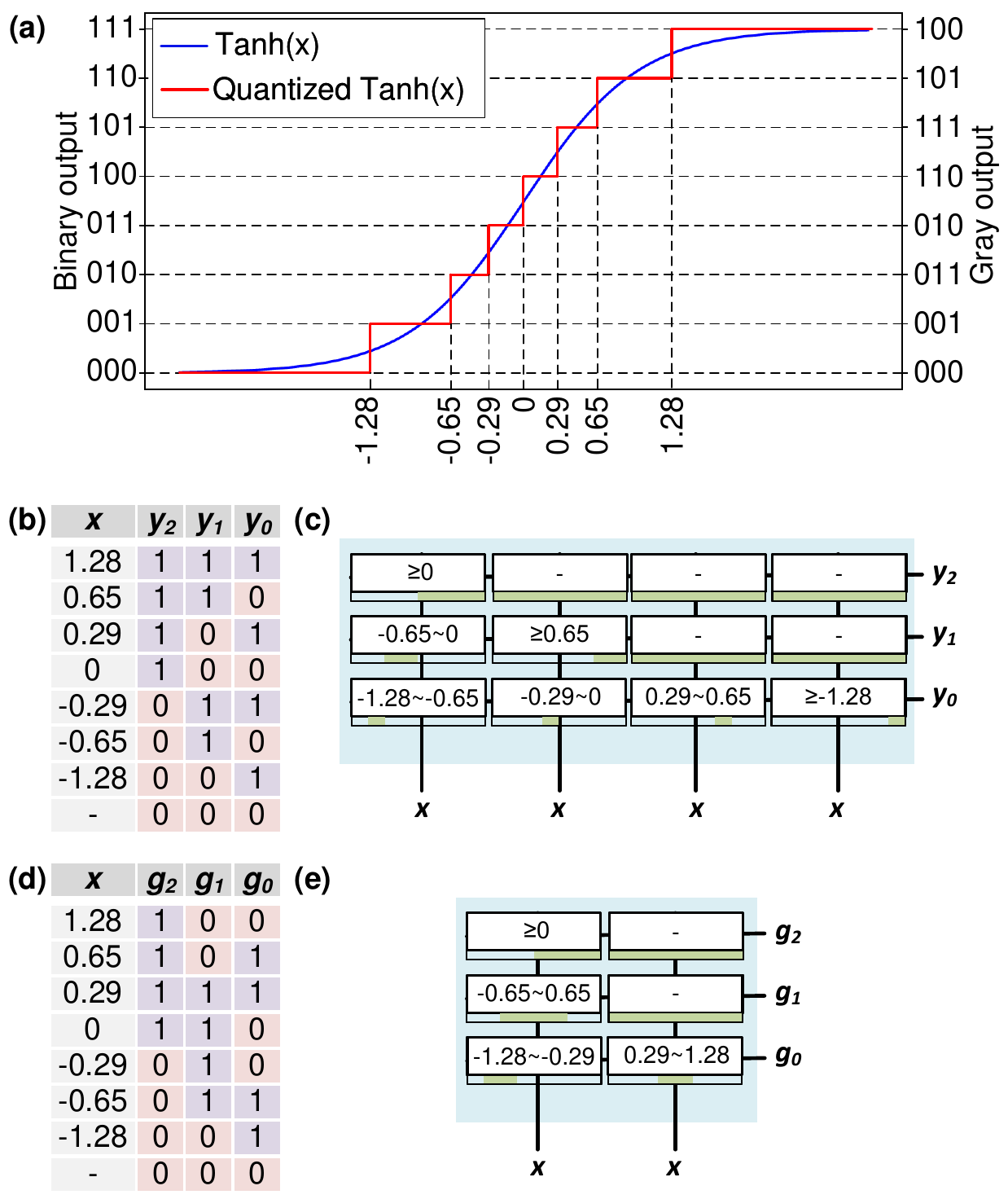}}
\caption{(a) Tanh activation output quantized to 3 bits in binary (left Y-axis) and Gray code (right Y-axis). (b) Binary output bits and corresponding Tanh inputs. (c) Mapping of (b) to a RACE array. (d) Gray code output bits and corresponding Tanh inputs. (e) Mapping of (d) to a RACE array.}
\label{fig:one-var-func}
\end{figure}

We use the Tanh function as an example to demonstrate how the RACE can compute arbitrary activation functions.
Figure~\ref{fig:one-var-func}(a) shows the Tanh curve in blue, along with a quantized version in red where the output $y$ is represented using 3 bits (left Y axis).
While in practice we use 8-bit quantization, here we show 3 bits for illustrative purposes.
Along the X-axis, we mark the input $x$ values whose Tanh outputs fall between two consecutive quantized levels.
From this graph, we derive a table shown in Figure~\ref{fig:one-var-func}(b), which lists the corresponding input ranges for each output bit.

By focusing on a specific output bit, we can identify the input ranges that result in that bit being set to 1.
For example, output bit $y_1$ is 1 when $-0.65 \leq x < 0$ or $x \geq 0.65$.
Figure~\ref{fig:one-var-func}(c) illustrates how these input ranges are mapped to a RACE array.
The RACE cells are configured in the \textit{one-variable 8-bit comparison mode}, with each row responsible for computing a specific output bit.
Cells marked with a dash are programmed to cover the full range and thus do not affect the result.

The number of rows in the RACE array corresponds to the number of output bits, while the number of columns is determined by the maximum number of disjoint input ranges across all bits.

\subsection{Data Dependent Multiplication}

\begin{figure*}[htbp]
\centerline{\includegraphics[width=7.3in]{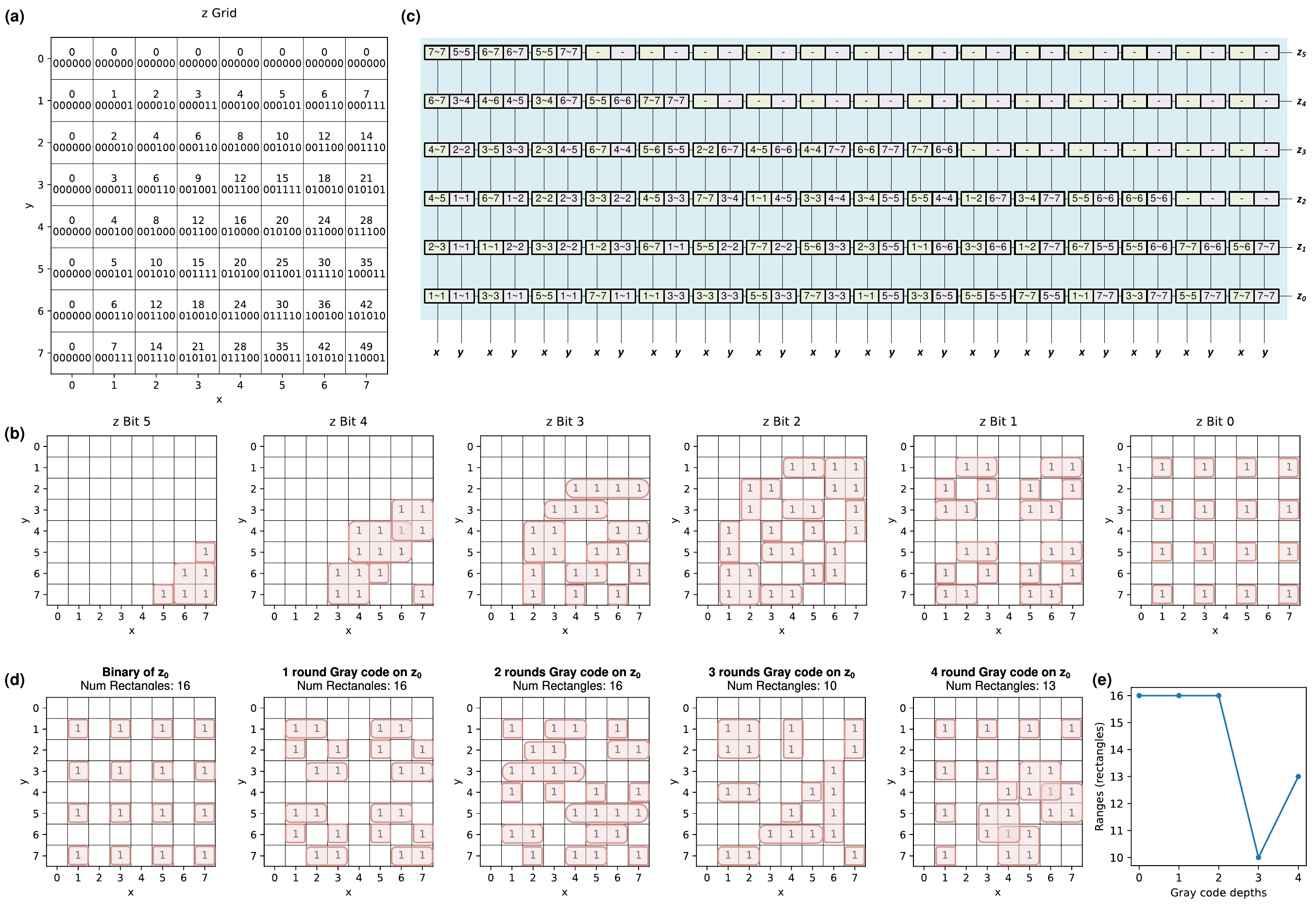}}
\caption{(a) Grid for 3-bit multiplication. (b) Grids per output bit. (c) Mapping (b) to a RACE array. (d) $z_0$ grid under different Gray code depths. (e) Number of ranges per grid in (d) for $z_0$ computation.}
\label{fig:two-var-func}
\end{figure*}

Figure~\ref{fig:two-var-func}(a) illustrates an example of 3-bit integer multiplication.
Although RACE supports 4-bit two-variable computations (and thus 4-bit multiplication), we use a 3-bit example here for clarity.\footnote{The attention mechanism in Transformers typically uses 8-bit multiplications, which can be implemented by combining four 4-bit multiplications implemented with RACE.}
The X- and Y-axes represent the value of two input operands, and each grid cell shows their 6-bit product in both decimal and binary formats.

Figure~\ref{fig:two-var-func}(b) separates the output into six individual grids—one for each bit of the 6-bit product.
Each grid resembles a Karnaugh map, where we identify rectangles that cover all cells with a value of 1.
Each rectangle corresponds to a specific input range combination of $x$ and $y$ that results in that output bit being set.
For instance, in the grid for $z_5$ (the leftmost grid in Figure~\ref{fig:two-var-func}(b)), at least three rectangles are needed to cover all the 1s.
This means $z_5 = 1$ when $x$ and $y$ fall within any of the following input combinations: (1) $x = 7$ and $y = 5$; (2) $x = 5$ and $y = 7$; (3) $6 \leq x \leq 7$ and $6 \leq y \leq 7$.

Figure~\ref{fig:two-var-func}(c) shows how input ranges are mapped to a RACE array configured in the \textit{two-variable 4-bit comparison mode}.
In this mode, each cell stores a range for both $x$ (green) and $y$ (purple).
Each row computes one output bit, and the number of columns is determined by the maximum number of rectangles (i.e., input range combinations) across all output bits.

\subsection{Softmax}

Softmax is a critical component in Transformer models and remains a challenging aspect for accelerators, as it involves costly exponentiation and division operations.
The Softmax function is defined as follows:
\begin{equation}
\text{softmax}(x_i)=\frac{e^{x_i}}{\sum_{_j=1}^{L}e^{x_j}}
\label{eq:safesoftmax}
\end{equation}
Here, $x_i$ represents an element in the input vector $\mathbf{x}=[x_1,...,x_L]$, and $L$ denotes the length of $\mathbf{x}$.

\begin{figure}[htbp]
\centerline{\includegraphics[width=3in]{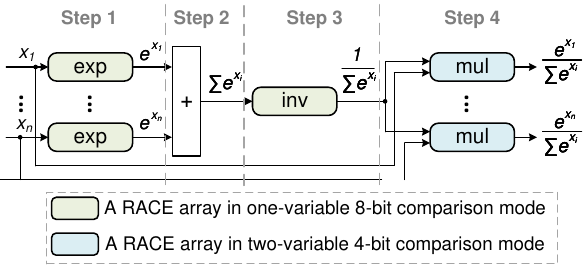}}
\caption{Dataflow of computing Softmax using RACE}
\label{fig:softmax}
\end{figure}

The expensive exponentiation and division operations can be easily mitigated with RACE-based computing.
Figure~\ref{fig:softmax} portrays the dataflow of computing Softmax with the RACE.
We first leverage the \textit{one-variable 8-bit comparison mode} for computing the exponentiations (step 1).
Subsequently, these exponents are aggregated by an adder tree to yield the denominator in Equation~\ref{eq:safesoftmax} (step 2).
To circumvent the division operation and its associated hardware costs, we replace the division with an inverse of the denominator followed by a multiplication:
\begin{equation}
a/b = a \cdot \frac{1}{b}
\end{equation}
The inverse can be implemented in the RACE with the \textit{one-variable 8-bit comparison mode} (step 3), and the multiplication can be implemented in the RACE with the \textit{two-variable 4-bit comparison mode} (step 4).

\section{Encoding}

Figures~\ref{fig:one-var-func}(c) and \ref{fig:two-var-func}(c) show many "Don't Care" cells — programmed with full range and not affecting results — due to varying input range requirements across output bits.
In the \textit{one-variable 8-bit comparison mode}, this depends on consecutive 1s per output bit in Figure~\ref{fig:one-var-func}(b).
In the \textit{two-variable 4-bit comparison mode}, it depends on the number of rectangles covering 1s in the Karnaugh-like grid in Figure~\ref{fig:two-var-func}(b).
Ultimately, it depends on how often each output bit toggles with input changes.

To reduce the RACE array size, we propose a Gray code-based optimization that lowers the number of input ranges.
Since Gray code ensures only one bit changes between consecutive values, it naturally reduces the bit toggling frequency.
Table~\ref{tab:gray} demonstrates an example of the Gray code format for 3-bit case.

\begin{table}[h!]
\centering
\caption{3-bit Gray Code Encoding.}
\label{tab:gray}
\begin{tabular}{|ccc|ccc|}
\hline
\textbf{Decimal} & \textbf{Binary} & \textbf{Gray} & \textbf{Decimal} & \textbf{Binary} & \textbf{Gray}\\
\hline
0 & 000 & 000 & 4 & 100 & 110 \\
1 & 001 & 001 & 5 & 101 & 111 \\
2 & 010 & 011 & 6 & 110 & 101 \\
3 & 011 & 010 & 7 & 111 & 100 \\
\hline
\end{tabular}
\vspace{-0.1in}
\end{table}

Instead of representing the outputs in binary format, we use Gray code.
The right Y-axis in Figure~\ref{fig:one-var-func}(a) shows the 3-bit Gray-coded Tanh output.
The corresponding input ranges for each output bit are listed in Figure~\ref{fig:one-var-func}(d).
This change reduces the number of input ranges for all bits except the MSB, cutting the RACE array size almost in half, as shown in Figure~\ref{fig:one-var-func}(e).

For the two-variable case, the effect is more subtle.
The first two grids in Figure~\ref{fig:two-var-func}(d) show that although Gray coding $z_0$ groups the 1s more tightly, it doesn't reduce the number of rectangles needed.
Since $z_0$ requires the most rectangles, this means no immediate savings in the RACE array size.
However, we notice that Gray-coded $z_0$ looks similar to binary $z_1$.
Applying Gray coding again clusters the 1s further, which may reduce the number of rectangles.
The remaining grids in Figure~\ref{fig:two-var-func}(d) show $z_0$ after additional rounds of Gray coding.
We observe a repeating pattern across encoding depths: the number of rectangles initially decreases and then increases again, as shown in Figure~\ref{fig:two-var-func}(e).

\begin{figure}[htbp]
\vspace{-0.1in}
\centerline{\includegraphics[width=2in]{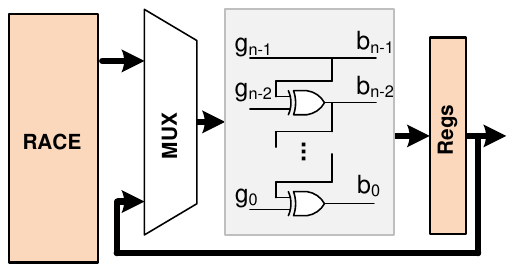}}
\caption{Decoding Gray code to binary format.}
\label{fig:decode}
\end{figure}

Since Gray code cannot be used directly in later computations, we need to convert it back to binary format. The conversion is straightforward: each binary bit is computed by XOR-ing all higher-order bits of the Gray code. The MSB remains unchanged between Gray and binary formats, i.e.,
\begin{equation}
b_i = \begin{cases}
g_i &i=n-1\\
XOR(g_{n-1}, g_{n-2}, ..., g_{i+1}) &i<n-1
\end{cases}
\end{equation}
where $b_i$ and $g_i$ are the $i$th bit in the binary and Gray code format, respectively.
$n$ is the bit width of the output.
Therefore, the conversion only needs some simple XOR gates, as shown in Figure~\ref{fig:decode}.
To support multi-depth Gray code decoding in the \textit{two-variable 4-bit comparison mode}, the outputs of the XOR gates are latched and fed back for additional decoding rounds.

\section{Analog circuit non-idealities}

\subsection{Circuit Simulation}
\label{sec:simulation}

\begin{figure}[htbp]
\vspace{-0.1in}
\centerline{\includegraphics[width=3in]{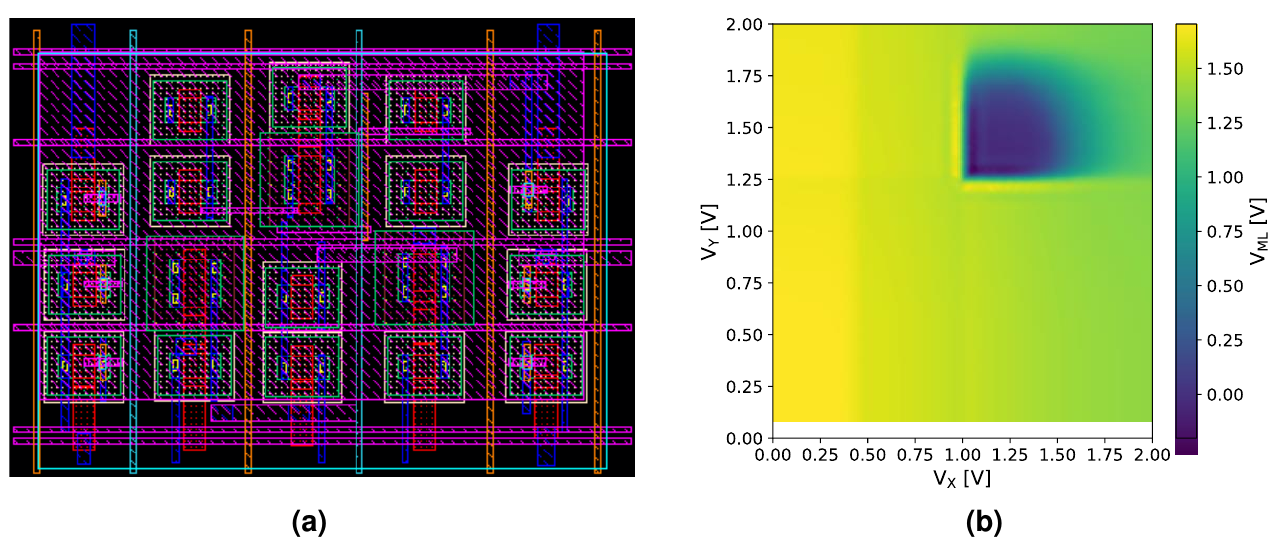}}
\caption{(a) Layout of a RACE cell in 22nm. (b) Computing a two-variable comparison in a RACE cell.}
\label{fig:circuit}
\end{figure}

Figure~\ref{fig:circuit}(a) shows the layout of an individual RACE cell designed using Global Foundries' GF22FDX 22 nm technology node.
Based on this design, we validate the cell’s functionality for two-variable comparison, as illustrated in Figure~\ref{fig:circuit}(b).
To simulate the input variations of $X$ and $Y$, we sweep the $DL$s accordingly (see Figure~\ref{fig:race}(c)).
The RRAM conductance values are programmed to represent the range condition $(4 < X < 9) \wedge (7 < Y < 12)$.
As shown in Figure~\ref{fig:circuit}(b), the $ML$ voltage drops within the specified input range and remains high outside of it, confirming correct comparison behavior.

\subsection{Nose-Aware Fine-Tuning}

\begin{algorithm}[tbp]
  \footnotesize
  \caption{Differentiable Approximation of two-variable comparison}
  \label{algo:race}
  \SetAlgoLined
  \KwIn{$x$, $y$: input values;
        $L_x$, $L_y$: lower bounds of $x$ and $y$; 
        $U_x$, $U_y$: upper bounds of $x$ and $y$;
        $\epsilon$: small number avoid dividing by 0;}
  \KwOut{$m$: one output bit of comparing $(L_x<x<U_x)\wedge(L_y<y<U_y)$}

$m_x = \mathrm{ReLU}(x - L_x) \cdot \mathrm{ReLU}(U_x - x)$\;
$m_y = \mathrm{ReLU}(y - L_y) \cdot \mathrm{ReLU}(U_y - y)$\;
$m = m_x + m_y$\;
$m = m / (m + \epsilon)$\;
\end{algorithm}

Analog noise affects both the crossbars and the RACE.
To mitigate crossbar noise, we use analog slicing~\cite{pedretti2021redundancy} when programming the weight matrix.
For noise in the RACE, we adopt a fine-tuning approach based on the ACAM noise model from~\cite{zhao2024noise}.
A key challenge is the non-differentiability of the comparison and Gray code decoding operations, which prevents gradient-based optimization.
To address this, we employ differentiable approximations~\cite{pedretti_differentiable_2022}.

While differentiable approximations for single-variable comparison and Gray decoding are available in~\cite{zhao2024noise}, we design a new differentiable formulation for the two-variable comparison, shown in Algorithm~\ref{algo:race}.
Each RACE cell originally computes $(L_x < x < U_x) \wedge (L_y < y < U_y)$.
We approximate these comparisons using subtraction and ReLU, replacing the logical AND with multiplication.
This yields a non-zero result only when all conditions are met.
Line 4 further scales the output to a value close to 0 or 1, mimicking the binary output of a RACE row.

These differentiable approximations enable end-to-end backpropagation, allowing us to fine-tune the bounds stored in the RACE cells and adapt the model to analog non-idealities.

\section{Methodology}

\begin{figure}[htbp]
\vspace{-0.1in}
\centerline{\includegraphics[width=3.5in]{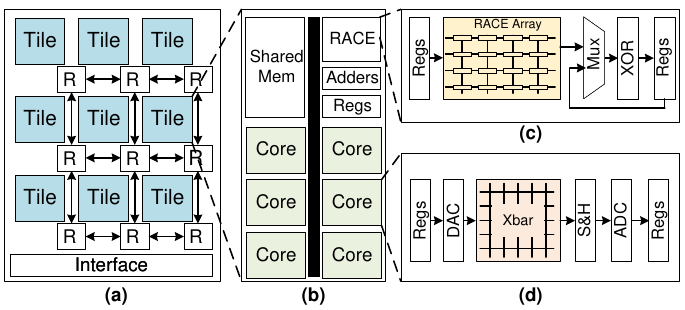}}
\caption{(a) A RACE-IT accelerator chip. (b) A tile in RACE-IT. (c) A RACE Unit. (d) A core in the tile.}
\vspace{-0.2in}
\label{fig:architecture}
\end{figure}

The overall architecture of our RACE-IT accelerator, which integrates the RACE into a crossbar-based IMC system, is shown in Figure~\ref{fig:architecture}.
The design largely follows ISAAC~\cite{shafiee2016isaac}, adopting a hierarchical structure of tiles and cores (Figure~\ref{fig:architecture}(a) and (b)), with the key difference that each tile integrates a RACE unit, adders, and registers to support intermediate data storage for RACE and adder operations.
Figure~\ref{fig:architecture}(c) illustrates the structure of a RACE unit, which consists of four RACE arrays, each sized $8 \times 50$ — determined in Section~\ref{sec:tradeoff}.
Each RACE unit also includes XOR-based decoding logic to convert Gray code outputs back to binary format.
We designed RACE-IT in 22nm CMOS technology with monolithically integrated TaOx RRAM~\cite{sheng2019lowconductance}.
%The results from Section~\ref{sec:simulation} is scaled to 22nm to design the RACE unit.
We considered conductance values from 0.1$\mu S$ to 150$\mu S$ both for the crossbar arrays and the RACE, and noise models are extracted from experimental data~\cite{sheng2019lowconductance}.

We compare RACE-IT against the following baselines: (1) NVIDIA H100 GPU, (2) Eyeriss~\cite{chen_eyeriss_2017}, (3) PUMA~\cite{ankit2019puma}, and (4) ReTransformer~\cite{yang2020retransformer}.
Performance and energy for the H100 GPU are measured using NVIDIA’s NVTX library with tensor cores enabled.
RACE-IT and all other baselines are implemented in Timeloop+Accelergy for evaluation.
We extend Eyeriss with the Flex-SFU~\cite{reggiani2023flex} to support non-VMM operations.
Since ReTransformer still uses digital logic for activation, we also equip it with Flex-SFUs for fair comparison.

We evaluate RACE-IT using BERT-base~\cite{devlin2019bert} on the GLUE benchmark to assess its support for core Transformer operations and the effectiveness of NAF. 
To demonstrate benefits in CNNs, we choose EfficientNet~\cite{tan2019efficientnet}, which uses non-ReLU activations unlike typical CNNs.
For scalability analysis, we also include LLaMA3.2~\cite{grattafiori2024llama} 1B and 3B models.

%%%%%%%%%%%%%%%%%%%%%%%%%%%%%%%%%%%%%%%%%%%%%%%%%%%%%%%%%%%%%%%%%%%%%%%%%%%%%%%%%%
%%%%%%%%%%%%%%%%%%%%%%%%%%%%%%%%%%%%%%%%%%%%%%%%%%%%%%%%%%%%%%%%%%%%%%%%%%%%%%%%%%
%%%%%%%%%%%%%%%%%%%%%%%%%%%%%%%%%%%%%%%%%%%%%%%%%%%%%%%%%%%%%%%%%%%%%%%%%%%%%%%%%%
\section{Experimental Results}

We first analyze the accuracy of RACE-IT-based computation, then explore the design space and associated trade-offs.
Based on these insights, we finalize the architecture configuration and thoroughly evaluate RACE-IT’s performance and energy-saving benefits.

\subsection{Accuracy}

\begin{table*}[ht]
\caption{Accuracy of various stages in the proposed NAF.}
\label{tbl:accuracy}
\begin{center}
\resizebox{\textwidth}{!}{%
\begin{tabular}{c | c | c c c c c c c c c}
\hline
\textbf{Model} & \textbf{EfficientNet} & \textbf{BERT-base} & \textbf{BERT-base} & \textbf{BERT-base} & \textbf{BERT-base} & \textbf{BERT-base} & \textbf{BERT-base} & \textbf{BERT-base} & \textbf{BERT-base} \\
Dataset & Cifar10 & cola* & mrpc & stsb$^+$ & rte & sst2 & qnli & mnli & qqp \\
\hline
\textbf{FP32}            & 91.15 & 0.585 & 84.55 & 0.88  & 64.98 & 92.43 & 91.54 & 84.2  & 90.92\\
\textbf{Crossbar noise}  & 91.14 & 0.585 & 84.55 & 0.88  & 65.7  & 92.43 & 91.46 & 84.02 & 90.9\\
\textbf{Quant}           & 89.32 & 0.591 & 84.31 & 0.873 & 63.17 & 91.74 & 89.3  & 81.99 & 90.3\\
\textbf{RACE}             & 89.47 & 0.578 & 84.31 & 0.872 & 64.25 & 91.85 & 89.27 & 82.23 & 90.17\\
\textbf{RACE noise}       & 70.25 & 0.515 & 83.22 & 0.85  & 62.05 & 91.36 & 87.65 & 77.42 & 89.27\\
\textbf{NAF}             & 89.37 & 0.577 & 84.21 & 0.87  & 64.81 & 91.69 & 89.21 & 82.09 & 90.11\\
\hline
\end{tabular}
}
\begin{flushleft}
\scriptsize{$^+$ = Pearson Correlation Coefficient (PCC), range [-1, 1], higher is better. \quad\quad  * =  Matthews Correlation Coefficient (MCC), range [-1, 1], higher is better.}\\
\end{flushleft}
\end{center}
\end{table*}

Table~\ref{tbl:accuracy} presents the accuracy results at various stages of applying NAF to CNN and Transformer models.
The \textbf{FP32} row serves as the baseline, showing the accuracy of the original models using 32-bit floating-point precision.
In the \textbf{Crossbar Noise} row, we simulate the effect of programming DNN weights onto RRAM crossbars using analog slicing.
Although analog noise is introduced to crossbars at this stage, the impact on accuracy is negligible thanks to the robustness provided by analog slicing.
The \textbf{Quant} row shows the accuracy after quantizing the outputs of activations, Softmax, and DMMuls into 8-bit fixed-point representations.
This quantization step is required because these operations will later be implemented with the RACE, which only produces 8-bit output.
Since weights are stored in RRAM crossbars, no additional quantization is needed for them, and overall accuracy remains largely unaffected.
In the \textbf{RACE} row, we replace the activations, Softmax, and DMMuls with RACE-based implementations.
Because RACE can accurately compute each output bit of these functions, there is no loss in accuracy.
However, when we further incorporate RRAM non-idealities into the RACE arrays, as shown in the \textbf{RACE noise} row, we observe a noticeable drop in accuracy due to analog variability and the non-linear comparison in RACE.
Finally, the last \textbf{NAF} row shows that the accuracy is recovered close to the ideal case in the \textbf{RACE} row after applying our proposed NAF.

\subsection{Design Space Exploration and Trade-offs}
\label{sec:tradeoff}

\begin{figure*}[htbp]
\centerline{\includegraphics[width=7in]{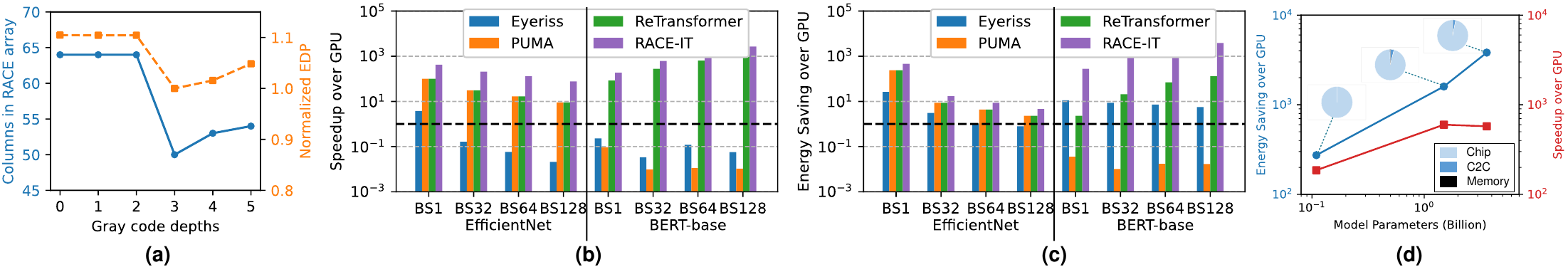}}
\caption{(a) RACE array size and EDP using binary and Gray code outputs at varying depths. (b) Speedup and (c) energy savings of RACE-IT and baseline accelerators normalized to GPUs. (d) Speedup and energy savings when scaling up RACE-IT for larger models.}
\vspace{-0.2in}
\label{fig:performace}
\end{figure*}

We first analyze the required array size for 4-bit multiplication under varying Gray code depths (blue line in Figure~\ref{fig:performace}(a)).
A depth of 0 corresponds to direct binary output.
The number of rows of a RACE array is fixed at 8 (for 8 output bits), while the number of columns depends on the number of ranges (i.e., rectangles in Figure~\ref{fig:two-var-func}(b)'s grids) among all the output bits.
As can be seen in the figure, a Gray code depth of 3 yields the smallest array size.

We then evaluate the Energy Delay Product (EDP, lower is better) across depths.
Although deeper encoding requires more decoding rounds, the overhead is minimal due to the simplicity of the XOR-based decoding logic.
Thus, EDP is mainly affected by additional energy caused by a larger array size.
As shown by the yellow line in Figure~\ref{fig:performace}(a), the smallest array size (at depth 3) also achieves the lowest EDP, making it the most energy-efficient configuration.
Therefore, we use a Gray code depth of 3 and set the size of each RACE array to $8\times 50$ in all subsequent experiments.

\subsection{Performance, Energy and Scalability}

Figure~\ref{fig:performace}(b) and (c) show the speedup and energy savings of RACE-IT and other accelerators relative to the GPU baseline.
For single-sample inference (BS = 1), which is critical for low-latency, energy-efficient edge AI inference, GPU performance suffers from low utilization~\cite{moon2024lpu}.
In contrast, RACE-IT achieves 453$\times$ speedup and 354$\times$ energy savings on average.
Its advantage over other IMC accelerators stems from replacing costly VFUs and natively supporting attention mechanisms.

We also evaluate the impact of larger batch sizes, for which GPUs are significantly optimized.
For CNNs, where the RACE mainly accelerates activation functions, increasing the batch size slightly reduces relative gains, but it still consistently outperforms all baselines.
For LLMs, the RACE is also used for accelerating Softmax and DMMul, leading to greater speedup and energy savings for large batch sizes.

As model size grows, a single chip cannot accommodate the entire network.
To avoid RRAM wear-out from time-multiplexing, we adopt a multi-chip setup.
BERT-base fits in one RACE-IT chip, while LLaMA3.2-1B and 3B require 3 and 8 chips, respectively.
We assume a conservative 10Gbps, 30pJ/bit chip-to-chip (C2C) interconnect.
Even under these worst-case conditions, RACE-IT achieves hundreds of times speedup and energy savings over GPU (Figure~\ref{fig:performace} (d)).

\section{Conclusion}
This paper presents RACE-IT, a reconfigurable analog engine for non-VMM operations in Transformer models.
By combining a novel RACE cell design with an encoding-based optimization and noise-aware fine-tuning, RACE-IT achieves compact area, high flexibility, and robust analog computation.
Extensive evaluations across CNNs and Transformer models demonstrate significant gains in performance and energy efficiency over state-of-the-art accelerators.
We believe RACE can open up novel possibilities for in-memory computing, enabling the computation of arbitrary functions in the analog domain.

%\bibliographystyle{IEEEtranS}
%\bibliography{refs}

\end{document}